\documentclass[prl,aps,twocolumn]{revtex4}
\usepackage{graphicx}
\usepackage{epstopdf}
\usepackage{dcolumn}
\usepackage{bm}
\usepackage{color}
\usepackage{ulem}
\usepackage{amsfonts,amsthm}
\usepackage{amsmath}
\usepackage{amssymb}
\usepackage{braket}
\begin{document}
\newcommand{\bR}{\mbox{\boldmath $R$}}
\newcommand{\Ha}{\mathcal{H}}
\newcommand{\mh}{\mathsf{h}}
\newcommand{\mA}{\mathsf{A}}
\newcommand{\mB}{\mathsf{B}}
\newcommand{\mC}{\mathsf{C}}
\newcommand{\mS}{\mathsf{S}}
\newcommand{\mU}{\mathsf{U}}
\newcommand{\mX}{\mathsf{X}}
\newcommand{\sP}{\mathcal{P}}
\newcommand{\sL}{\mathcal{L}}
\newcommand{\sO}{\mathcal{O}}
\newcommand{\la}{\langle}
\newcommand{\ra}{\rangle}
\newcommand{\ga}{\alpha}
\newcommand{\gb}{\beta}
\newcommand{\gc}{\gamma}
\newcommand{\gs}{\sigma}
\newcommand{\vk}{{\bm{k}}}
\newcommand{\vq}{{\bm{q}}}
\newcommand{\vR}{{\bm{R}}}
\newcommand{\vQ}{{\bm{Q}}}
\newcommand{\vga}{{\bm{\alpha}}}
\newcommand{\vgc}{{\bm{\gamma}}}
\newcommand{\mb}[1]{\mathbf{#1}}
\def\vec#1{\boldsymbol #1}
\arraycolsep=0.0em
\newcommand{\Ns}{N_{\text{s}}}
%

\title{
Electronic Correlation and Geometrical Frustration in 
Molecular {Solids} \\ 
-- A Systematic {\it ab initio} Study of {$\beta^\prime$}-$X$[Pd(dmit)$_{2}$]$_{2}$
}

\author{
Takahiro Misawa$^{1,\dagger}$, Kazuyoshi Yoshimi$^{1,\dagger}$ and Takao Tsumuraya$^{2}$
}

\affiliation{$^1$Institute for Solid State Physics,~University of Tokyo,~5-1-5 Kashiwanoha, Kashiwa, Chiba 277-8581, Japan}
\affiliation{$^2$Priority Organization for Innovation and Excellence, Kumamoto University, 2-39-1 Kurokami, Kumamoto 860-8555, Japan}
\affiliation{\rm $^{\dagger}$The authors contributed to this work equally.}
\date{\today}

\begin{abstract}
We systematically derive low-energy effective Hamiltonians
for {molecular solids} 
{$\beta^\prime$}-$X$[Pd(dmit)$_{2}$]$_{2}$ ($X$ represents a cation)
using {\it ab initio} {density functional theory} calculations 
and clarify how the cation controls the {inter-dimer} 
transfer integrals and the interaction parameters.
The effective models are solved using the exact diagonalization method and 
the antiferromagnetic ordered moment is shown to be significantly
suppressed {around the spin-liquid candidate of $X$=EtMe$_{3}$Sb}, which is reported in experiments.
We also show that both the geometrical frustration 
and the off-site interactions play 
essential roles in the suppression of antiferromagnetic {ordering}.
This systematic derivation and analysis
of the low-energy effective Hamiltonians offer a firm basis
to clarify the nature of the 
quantum spin liquid found 
in {$\beta^\prime$}-EtMe$_{3}$Sb[Pd(dmit)$_{2}$]$_{2}$.
\end{abstract}


\maketitle

{\it Introduction---.}~Quantum spin liquids {(QSLs)}~\cite{Balents_Nature2010}, which are {Mott} insulators 
without any broken symmetry, even at zero temperature,
are new states of matter that have attracted
much interest in the last few decades. 
In pioneering work by Anderson and Fazekas~\cite{Anderson_MRB1973,Fazekas_PM1974},
it was proposed that geometrical frustration in the
magnetic interactions 
can melt 
magnetic order 
and induce a QSL. 
Motivated by such a proposal,
much theoretical and experimental work has been done
to search for QSLs
in frustrated magnetic materials~\cite{Balents_Nature2010,Savary_RP2016,Diep}.

{In the {molecular solids}, the} 
van der Waals interactions among molecules tend to align 
them in a closed-packed way; therefore, 
they often form frustrated lattices such as an anisotropic 
triangular lattice~\cite{Kanoda_Kato_2011ARCMP,Powell_RPP2011}. 
{Due to this feature, several families of {molecular solids} 
offer promising platforms for realizing QSLs, such as $\kappa$-(BEDT-TTF)$_{2}$$X$ and 
$\beta^{\prime}$-$X$[Pd(dmit)$_{2}$]$_{2}$, 
where $X$ represents anion and cation species, respectively 
[BEDT-TTF = bis(ethylenedithio)tetrathiafulvalene, dmit = 1,3-dithiole-2-thione-4,5-dithiolate)].}
It was reported that no clear symmetry breaking 
in $\kappa$-(BEDT-TTF)$_{2}$Cu$_2$(CN)$_3$
occurred down to 32 mK~\cite{Shimizu_RPL2003}, which
{indicates} that a QSL
due to geometrical frustration 
realizes in this compound.
QSLs have been discovered 
in several {other {molecular solids}} 
such as $\beta^{\prime}$-EtMe$_{3}$Sb[Pd(dmit)$_{2}$]$_{2}$ (Et = C$_2$H$_5$, Me = CH$_3$)~\cite{Tamura_JPhys2002,Itou_NPhys2010}
and $\kappa$-H$_{3}$(Cat-EDT-TTF)$_{2}$, 
where Cat-EDT-TTF is ethylenedithio-tetrathiafulvalene~\cite{Isono_PRL2014}.

Among several QSLs found in {molecular solids},
Pd(dmit)$_{2}$ salts offer an ideal platform for examining
{the key parameters that induce the QSL}
because it is possible to tune the ground states 
from the ordered states (antiferromagnetic ordering or charge ordering) to
a QSL by systematically changing the cations~\cite{Kanoda_Kato_2011ARCMP}. 
{
In addition to that, because the high-quality samples with less impurities
are available, detailed measurement of the thermal transport
was done to clarify the nature of the QSL~\cite{M_Yamashita_dmit10} and
it was proposed that the large thermal conductivity indicates 
the emergence of the exotic particle such as the spinon in the QSL.
However, this result has been challenged by the recent experiments
and the existence/absence of the exotic particle in the QSL 
is under hot debate~\cite{BH_PRX2019,Ni_PRL2019,M_Yamashita_JPSJ2019,M_Yamashita_PRB2020}.
Although the theoretical studies such as establishing the 
low-energy Hamiltonians are expected to play 
important role for resolving the contradiction,
there are a few theoretical studies based on non-empirical methods.
}

In previous studies,
the half-filled Hubbard model on an anisotropic triangular lattice 
has been 
obtained as an effective microscopic Hamiltonian to describe  
the electronic structures in Pd(dmit)$_2$ salts
~\cite{Nakamura,Powell_PRL12,Jacko_PRB2013} because
the bands crossing the Fermi level are half-filled and 
isolated from the other bands. 
These bands mainly originate from the antibonding pair of the
highest occupied molecular orbitals (HOMO) of the two Pd(dmit)$_2$ 
molecules that form a dimer~\cite{Kanoda_Kato_2011ARCMP, Kato_Chem_Rev, Miyazaki03_dmit, Miyazaki99_dmit}. 
A one band model with three interdimer transfer integrals 
reproduces the {density functional theory (DFT)} bands well ~\cite{Nakamura, Tsumu_Pd_dmit2_13}.
Through extended H\"{u}ckel calculations and 
tight-binding fitting {to DFT bands}, 
the amplitudes of the geometrical frustration, i.e.,
the anisotropy of the transfer integrals, have been evaluated.
It has thus been proposed that the 
geometrical frustration governs 
the magnetic properties of the
dmit salts and that the magnetic order can be suppressed 
by changing the transfer integrals~\cite{Kanoda_Kato_2011ARCMP}. 

In contrast to the transfer integrals,
information on the interaction parameters is 
limited, even though they play an important role 
in the stabilization and suppression of the magnetic order.
Although the interaction parameters have been evaluated 
only for the QSL compound $X=$ EtMe$_3$Sb~\cite{Nakamura} {in the studies of the dmit salts},
the compound dependence and role in inducing the 
QSL state have yet to be clarified.
{We note that 
previous studies are limited to derive 
the low-energy effective Hamiltonians and
it is not examined whether the 
low-energy effective Hamiltonians reproduce the cation dependence
of the ground states in Pd(dmit)$_2$ salts by solving them.}

{In this Letter,}
{to clarify the microscopic origin of the cation dependence of the
ground states}
in Pd(dmit)$_2$ salts,
we perform systematic $ab$ $initio$ derivations
of the low-energy effective Hamiltonians, 
including both the transfer integrals and 
the interaction parameters
for 
$\beta^{\prime}$-$X$[Pd(dmit)$_{2}$]$_{2}$.
At present, 9 different Pd(dmit)$_2$ salts {with $\beta^{\prime}$-type structure} are synthesized 
because the monovalent cations $X$ can take three different types, 
i.e., Me$Y_{4}$, EtMe$_2$$Y_{2}$, and EtMe$_3$$Y$,
where the choice of pnictogen $Y$ is P, As, and Sb.
Through comparison with the obtained 
low-energy effective models,
we find two trends in the parameters of the effective Hamiltonians:
{\bf 1.} Hopping parameters $t_{c}$/$t_{a}$ (definitions are given in Fig. 1(a)) 
increase in the order of P, As, Sb.
{\bf 2.} Onsite Coulomb interactions $U/t_{a}$ 
increase in the order of P, As, Sb (EtMe$_{3}$Sb is exceptional).

Furthermore, to identify how
the microscopic parameters affect the magnetic properties,
the obtained models are solved using exact diagonalization~\cite{hphi},
{which enable us to clarify the overall trend of the
magnetic properties of the frustrated magnets~\cite{Dagotto_1989PRB,Poilblanc_1991PRB}.}
As a result,
the trend in the compound dependence 
of the magnetic ordered moment is successfully reproduced
and the magnetic ordered moment is found to be significantly suppressed in
$X=$EtMe$_3$Sb, wherein the QSL state was reported experimentally.   
EtMe$_{3}$Sb is sandwiched between the Neel-type
antiferromagnetic order and the striped antiferromagnetic order when
the possibility of charge ordering is ignored.
We also show that both 
geometrical frustration and 
off-site interactions play key roles in suppression of the
antiferromagnetic order.
The present results clarify the microscopic origin of the 
QSL 
and offer a firm basis to 
comprehensively understand 
QSLs found in {molecular solids}.

{\it Ab initio derivation of effective models---.}
Based on $ab$ $initio$ calculations,
the following
single-band extended Hubbard-type Hamiltonian was obtained.
\begin{align}
&H=
\sum_{ij,\sigma}t_{ij}(c_{i\sigma}^{\dagger}c_{j\sigma}+{\rm h.c.})
+U\sum_{i}n_{i\uparrow}n_{i\downarrow}
+\sum_{ij}V_{ij}N_{i}N_{j} \notag \\
&+\sum_{ij,\sigma\rho}J_{ij}(c_{i\sigma}^{\dagger}c_{j\rho}^{\dagger}c_{i\rho}c_{j\sigma}
+c_{i\sigma}^{\dagger}c_{i\rho}^{\dagger}c_{j\rho}c_{i\sigma}),
\label{Ham}
\end{align}
where $c^{\dagger}_{i\sigma}$ ($c_{i\sigma}$) is a 
creation (annihilation) operator of an electron with spin $\sigma$
in the Wannier orbital localized at the $i$th dmit dimers.
The number operators are defined as
$n_{i\sigma}=c_{i\sigma}^{\dagger}c_{i\sigma}$ and 
$N_{i}=n_{i\uparrow}+n_{i\downarrow}$.
We evaluate the transfer integrals $t_{ij}$,
the on-site Coulomb interaction $U$,
the off-site Coulomb interaction $V_{ij}$,
and {direct} exchange interactions $J_{ij}$ in an $ab$ $initio$ way.
{
We note that the double-counting problem on the Hartree terms for the multi-orbital systems~\cite{Misawa_JPSJ2011,Seo_JPSJ2013}  
does not exist in this study because the employed model is the single-band model.}
From here, we detail how to evaluate these parameters. 

\begin{figure}[t!]
  \begin{center}
    \includegraphics[width=9cm]{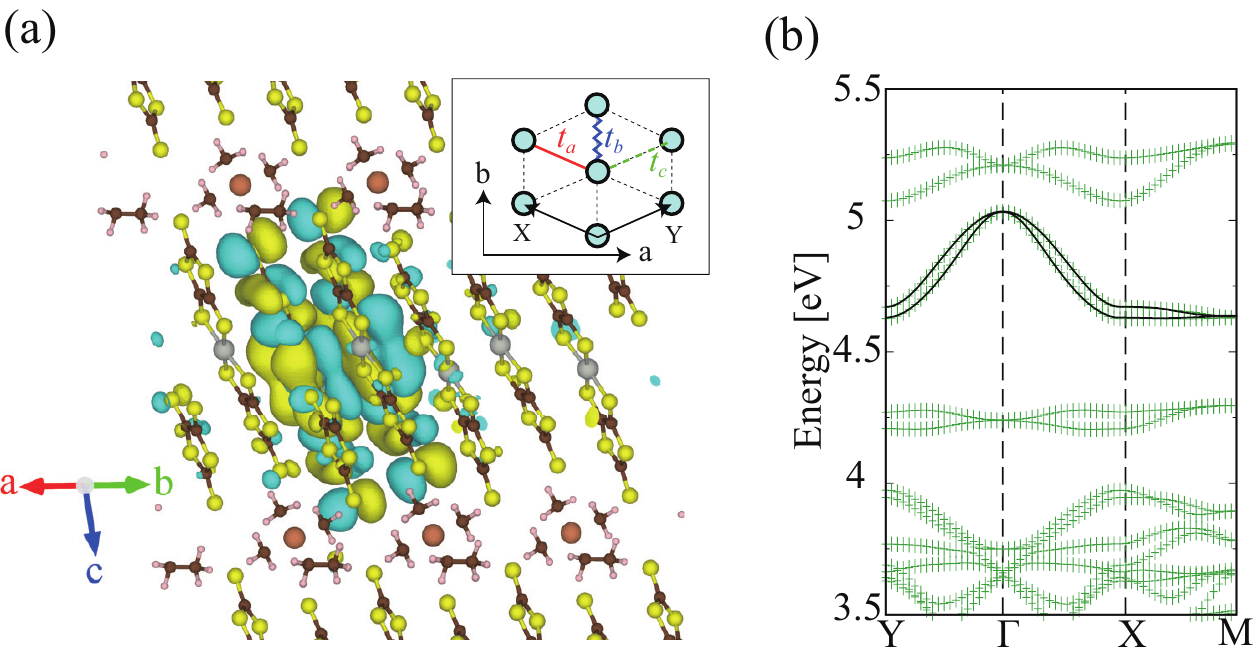}
  \end{center}
\caption{(color online)~
(a) Wannier function for EtMe$_3$Sb drawn using \texttt{VESTA}~\cite{VESTA}.
The inset shows a schematic of the lattice structure of EtMe$_3$Sb in the a-b plane. 
Circles correspond to Wannier centers and transfer 
integrals between the nearest-neighbor Wannier orbitals are shown.
(b) Band dispersion for EtMe$_3$Sb. The dotted (solid) lines are obtained by the first-principles DFT method (Wannier interpolation).}
\label{fig1}
\end{figure}

To derive low-energy effective Hamiltonians, 
we first perform non-spin-polarized calculations with a first-principles DFT method~\cite{H-K_1964, Kohn_Sham} and obtain the Bloch functions for experimental structures of the nine members of $\beta^{\prime}$-type $X$[Pd(dmit)$_2$]$_2$, of which the structures are measured above the antiferromagnetic or charge ordering transition temperature~\cite{Crystals12_Kato, KUeda_EtMe3Sb2018}. 
All the structures have the same space group symmetry of $C$2/$c$.
The positions of all the H atoms were relaxed because those determined 
from x-ray diffraction measurements typically show slightly shorter C--H bond distances than the DFT optimized positions.
One-electron Kohn-Sham equations are solved self-consistently 
using a pseudopotential technique with plane wave basis sets, 
which is implemented in the scalar relativistic code 
of \texttt{Quantum Espresso 6.3}~\cite{QE}.  
The exchange-correlation functional used is the generalized 
gradient approximation (GGA) proposed 
by Perdew, Burke, and Ernzerhof (PBE)~\cite{GGA_PBE}. 
The cutoff energies for plane waves and charge 
densities were set to be 70 and 280 Ry, respectively. 
A $5\times5\times3$ uniform k-point mesh was used with 
a Gaussian smearing method during self-consistent loops. 

After obtaining the Bloch functions, the
maximally localized Wannier functions (MLWF) were constructed using 
\texttt{RESPACK}~\cite{RESPACK}. 
To make the MLWF, the 
half-filled bands crossing the Fermi level were selected as
the low-energy degrees of freedom.
Initial coordinates of the MLWF 
were set to be at the center between two [Pd(dmit)$_2$] monomers to generate 
a one band model (so-called dimer model)~\cite{KinoFukuyama_dimer96}. 

In Fig. \ref{fig1}(a), we show the MLWF for $X$=EtMe$_3$Sb. 
The MLWF, of which the center position is located
at the center between two [Pd(dmit)$_2$] monomers, spreads over the molecule 
and forms the dimer unit. 
Figure \ref{fig1}(b) shows that the MLWF 
{nearly}
perfectly reproduces the band structures obtained by the DFT calculations.
Using the MLWF,
the transfer integrals of the low-energy 
effective models were evaluated as
\begin{align}
t_{nm}(\vec{R})=\langle\phi_{n,0}|H_{k}|\phi_{m,\vec{R}} \rangle,
\end{align}
where $\phi_{n,\vec{R}}$ is the $n$th  MLWF centered at $\vec{R}$,
and $H_k$ is the one-body part of the $ab$ $initio$ Hamiltonian.
The lattice structure in the dimer units is shown 
in the inset of Fig. \ref{fig1}(a). 
We note that $t_{a}$ is the largest transfer integral and
$t_{c}/t_{a}$ denotes the amplitude of the geometrical frustration 
as discussed later.

\begin{figure}[t!]
  \begin{center}
    \includegraphics[width=8cm]{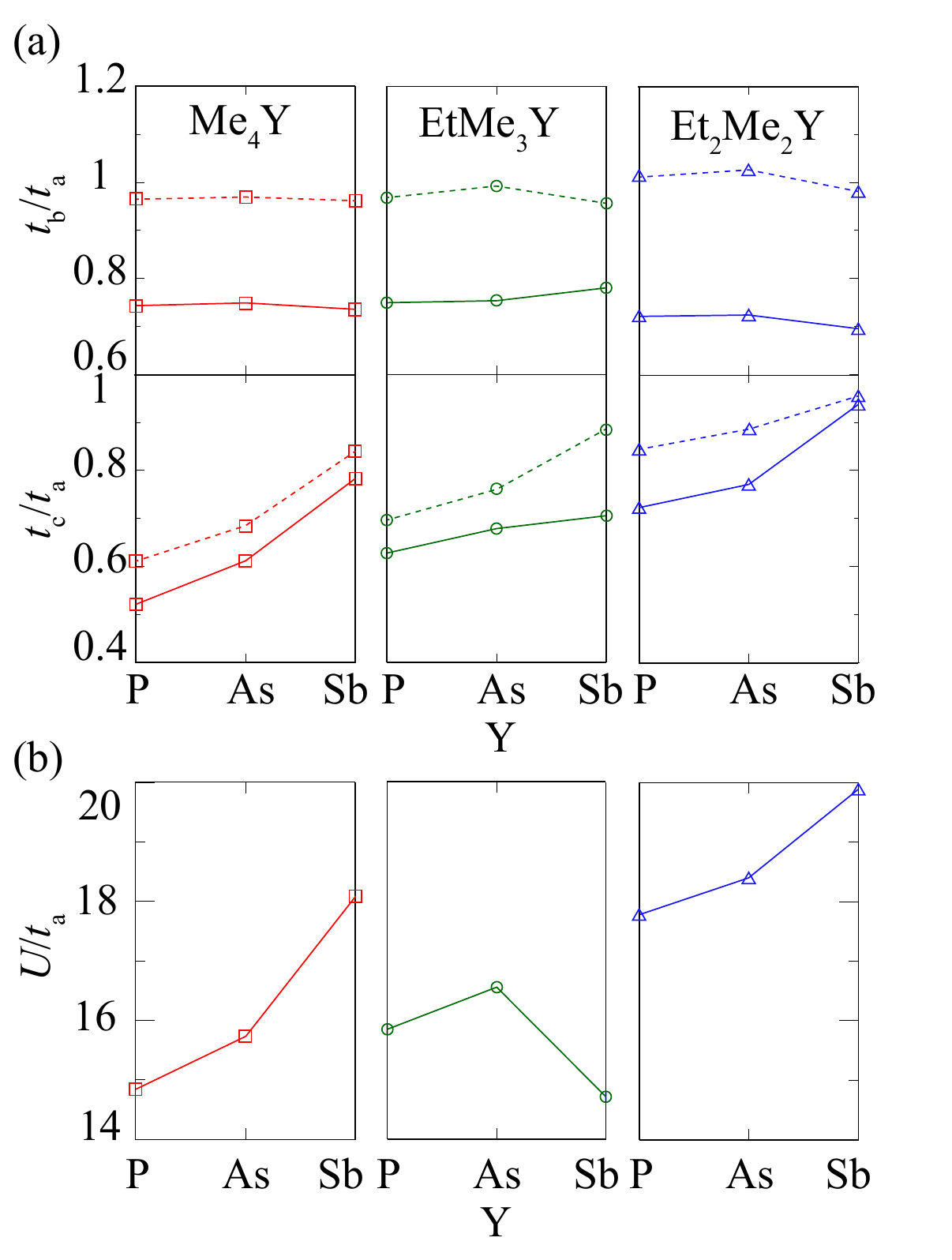}
  \end{center}
\caption{(color online)~
(a) Band dispersion for EtMe$_3$Sb. The dotted (solid) lines are obtained by the first-principles DFT method (Wannier interpolation). (b) Material dependency of $t_b/t_a$ and $t_c/t_a$. The dotted (solid) lines are obtained by the transfer integrals calculated by the extended H\"{u}ckel method \cite{Crystals12_Kato} (Wannier function basis).}
\label{fig2}
\end{figure}

Figure \ref{fig2}(a) shows the compound dependence of 
the normalized transfer 
integrals obtained by the MLWF fitting.
For comparison, the transfer integrals obtained by
the extended H\"{u}ckel calculations are shown.
Although the transfer integrals evaluated by the MLWF are slightly 
different from those obtained by the extended H\"{u}ckel calculations,
the trend of the compound dependence is consistent, i.e.,
$t_{b}/t_{a}$ is not largely dependent on the cations and
$t_{c}/t_{a}$ increases in the order of P, As, and Sb.
This trend is also found in the literature and the 
origin can be attributed to the {cation radius, which controls the
{distortion}
of the dmit molecules~\cite{Crystals12_Kato}.} 

The interactions were also evaluated by the constrained 
random-phase approximation (cRPA)~\cite{Aryasetiawan_PRB2004} method using \texttt{RESPACK}. 
The energy cutoff for the dielectric function was set to be 3 Ry. 
The interaction terms are given as follows:
\begin{align}
W_{nm,kl}(\vec{R}_{1},\vec{R}_{2},\vec{R}_{3},\vec{R}_{4})=
\langle\phi_{n\vec{R}_1}\phi_{m\vec{R}_2}| H_W|\phi_{k\vec{R}_3}\phi_{l\vec{R}_4}\rangle,
\end{align}
where $H_{W}$ represents the
interaction term of the $ab$ $initio$ Hamiltonians.
Although three-body and four-body interactions generally occur,
we only treat the two-body interactions, 
such as density-density interactions {$U_{mn}(\vec{R})= W_{mm,nn}(\vec{0},\vec{0},\vec{R},\vec{R})$}
(that is, the on-site and off-site Coulomb interactions)
and the {direct} exchange interactions {$J_{mn}(\vec{R})= W_{mn,nm}(\vec{0},\vec{R},\vec{R},\vec{0})$}
because the amplitudes of other terms are negligibly small.
{In 
\cite{SM}, we show number of the screening bands dependence
of the interaction parameters. We note that 
when the number of screening bands becomes large,
the cRPA method can be valid in contrast 
to the previous studies~\cite{Shinaoka_PRB2015,Honerkamp_PRB2018}.}

Figure \ref{fig2}(b) shows the compound dependence of 
the normalized onsite Coulomb interaction $U/t_{a}$.
$U/t_{a}$ increases in the order of P, As, Sb, as in the
transfer integrals, except for EtMe$_3$Sb.
{As we have detailed in \cite{SM}, the compound dependence of $U/t_{a}$ is mainly
controlled by changes in $t_{a}$
because $U$ is not largely dependent on the compounds. 
{The transfer integral $t_{a}$ 
decreases in the order of P, As, and Sb because the 
distortions of the molecules become larger 
in this order~\cite{Crystals12_Kato}.}
In contrast, $U$ does not show a large compound dependence (see S.1 in \cite{SM}),
and for this reason, $U/t_{a}$ increases in the order of P, As, and Sb.
Low-temperature structures were employed for EtMe$_3$Sb; therefore,
$t_{a}$ becomes large and $U/t_{a}$ becomes small. 
This is the origin of the 
exceptional behavior observed in EtMe$_{3}$Sb.}
{We also point out that $U/W$ ($W$ is the bant width) is estimated
as $U/W\sim 2.0$ for EtMe$_{3}$Sb, which is roughly
consistent with the experimental estimation ($U/W\sim 2.3$)~\cite{Pustogow_2018NM}.}

{\it Analysis of effective models---}
To clarify how the geometrical frustration
and the interaction parameters affect the 
magnetic properties in the low-energy
effective models defined in Eq.~(\ref{Ham}), 
exact diagonalization was performed for
small clusters (system size is $N_{s}=4\times4$ with 
the periodic boundary conditions).
In the calculations, we take the transfer integrals,
off-site Coulomb interactions, and the {direct} exchange interactions
up to the next-nearest neighbor.
Using exact diagonalization,
the compound dependence of the magnetic properties can be clarified
without relying on any specific approximations.
It should be noted that it is necessary to introduce
a constant shift in the interaction parameters to reflect
the two-dimensionality of the effective models, and to obtain
physically reasonable results.
Following the parameter-free approach, we 
employ a constant shift of $\Delta=0.30$ eV for all compounds, 
which is a value comparable to that in the literature~\cite{Nakamura}.
In S.~2~\cite{SM}, we examine the
effects of $\Delta$ and confirm that
the changes in the constant shift 
do not change the results significantly. 

\begin{figure}[htb!]
  \begin{center}
   \includegraphics[width=9cm,clip]{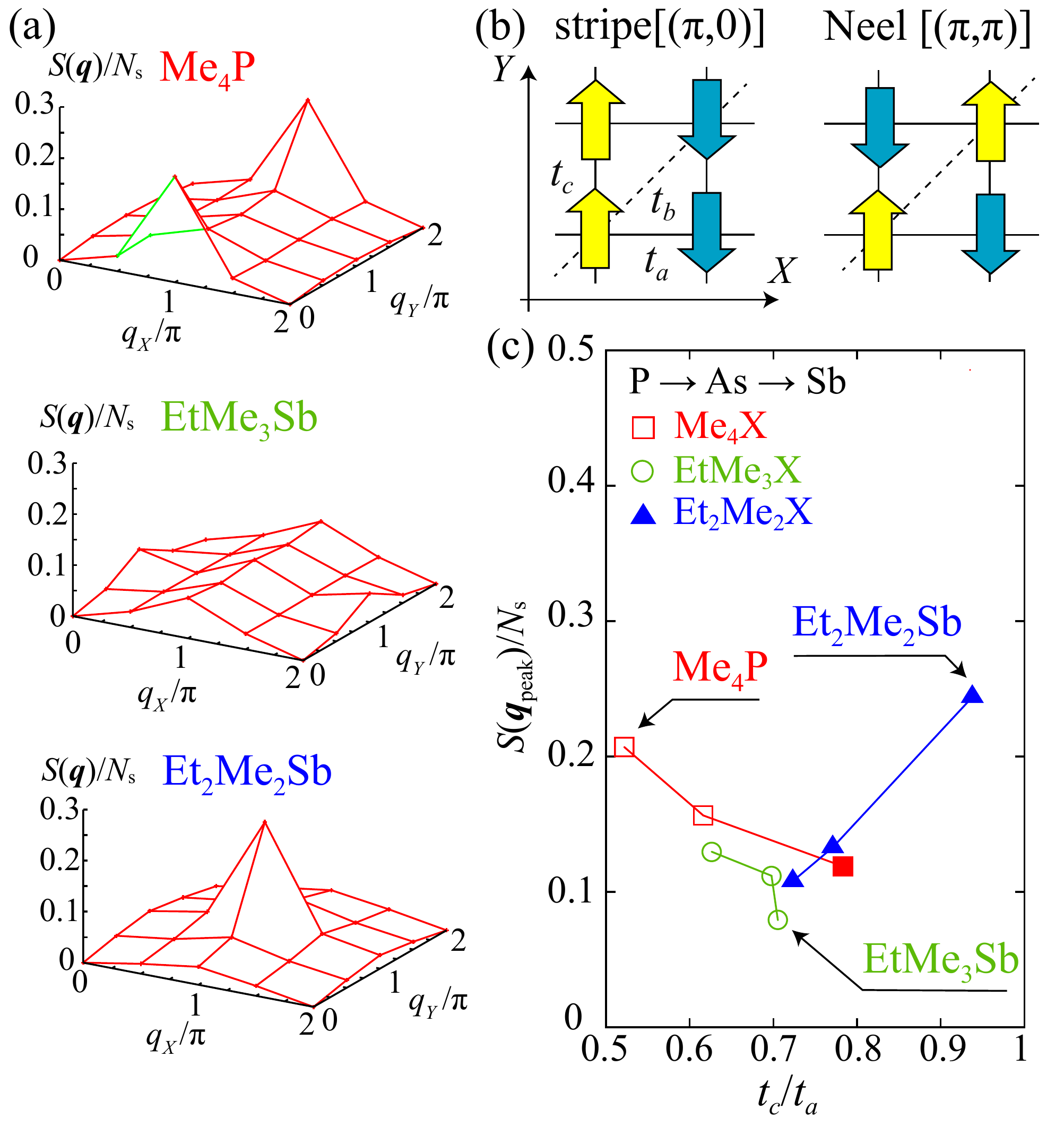}
  \end{center}
\caption{(color online)~
(a)~Spin structure factors for three typical compounds.
A significant reduction of the Bragg peak occurs in EtMe$_3$Sb.
(b)~Schematic diagrams for stripe and Neel ordered states.
(c)~ Dependence of the spin structure factors on $t_{c}/t_{a}$.
For each family, $t_{c}/t_{a}$ 
increases in the order of P, As, and Sb.
}
\label{fig3}
\end{figure}

Figure ~\ref{fig3}(a) shows the 
spin structure factors 
\begin{align}
S(\vec{q})=\frac{1}{N_{s}}\sum_{i,j}
\langle\vec{S}_{i}\cdot\vec{S}_{j}\rangle
e^{i\vec{q}(\vec{r}_{i}-\vec{r}_{j})}
\end{align}
for several compounds, i.e.,
$X$=Me$_4$P, EtMe$_3$Sb, and Et$_2$Me$_2$Sb 
(the spin structure factors for all nine compounds are shown in S.~3~\cite{SM}.).
In Me$_4$P, which shows the highest Neel temperature and anisotropy of transfer integrals, the spin structure factor is a sharp peak at $\vec{q}=(\pi,0)$, which indicates 
a stripe magnetic order (top panel in Fig. ~\ref{fig3}(b)).
The amplitudes of the spin structure factors
are significantly suppressed in EtMe$_3$Sb and
no clear signature of magnetic order is evident.
{Additionally, we note that no clear signatures of 
other exotic non-magnetic phases, such as the bond-order phase~~\cite{MNakamura_JPSJ1999,MNakamura_PRB2000},
are observed (see S. 4 in \cite{SM}).}
This indicates that the  
QSL state is formed in this compound.
On the other hand,
for Et$_2$Me$_2$Sb, the sharp peak
appears at $\vec{q}=(\pi,\pi)$, which indicates a Neel-type
magnetic order shown in the bottom panel in Fig. ~\ref{fig3}(b).
The ground state is a charge ordered state \cite{Nakao_JPSJ_Et2Me2Sb}; 
therefore, this result is apparently
inconsistent with previous experimental results.
However, this discrepancy can be attributed to the 
discarding of the long-range part of the Coulomb interactions;
in a 4$\times$4 systems size, only the
the Coulomb interactions up to the next-nearest neighbor can be treated, and
the charge-order pattern cannot be accurately determined.
{We note that, it is pointed out the
the interplay of the long-range Coulomb interactions and
the electron-phonon couplings stalibilizes 
the charge-ordered phase in Et$_2$Me$_2$Sb~\cite{Seo_2015JPSJ}.}
Our analysis indicates that
Et$_2$Me$_2$Sb has antiferromagnetic instability
toward $(\pi,\pi)$, if we ignore 
the instability toward charge ordering.

Figure ~\ref{fig3}(c) 
shows the compound dependence of the 
peak values of the spin structure factor $S(\vec{q}_{\rm peak})$
as a function of $t_{c}/t_{a}$.
$X=$ EtMe$_{3}$Sb is located around the boundary of the
stripe and Neel magnetic order, and 
$S(\vec{q}_{\rm peak})$ is significantly
reduced at $X= $EtMe$_{3}$Sb.
The overall trend of the compound dependence of the spin structure
factors is consistent with
the experimental results~\cite{Kanoda_Kato_2011ARCMP}. 
It is especially noteworthy that the 
low-energy effective Hamiltonian
for EtMe$_{3}$Sb shows suppression of the 
spin structure factors, which is 
consistent with experimentally observed QSL behavior.

To clarify the origin of the 
magnetic ordered moment reduction, we systematically changed
the parameters in the $ab$ $initio$ effective Hamiltonian for EtMe$_3$Sb.
First, the effects of the 
geometrical frustration were examined by changing $t_{c}/t_{a}$.
In Fig. 4(a), we show the $t_{c}/t_{a}$ dependence of 
$S(\vec{q}_{\rm peak})$. By artificially decreasing $t_{c}/t_{a}$,
a sudden change in $S(\vec{q}_{\rm peak})$ occurs
at $t_{c}/t_{a}\sim0.58$, which 
indicates a first-order phase transition
between the stripe magnetic ordered phase 
and the possible QSL state.
This result clearly shows that the geometrical frustration 
plays a key role in suppression of the magnetic ordered moment.

\begin{figure}[bt!]
  \begin{center}
    \includegraphics[width=9cm,clip]{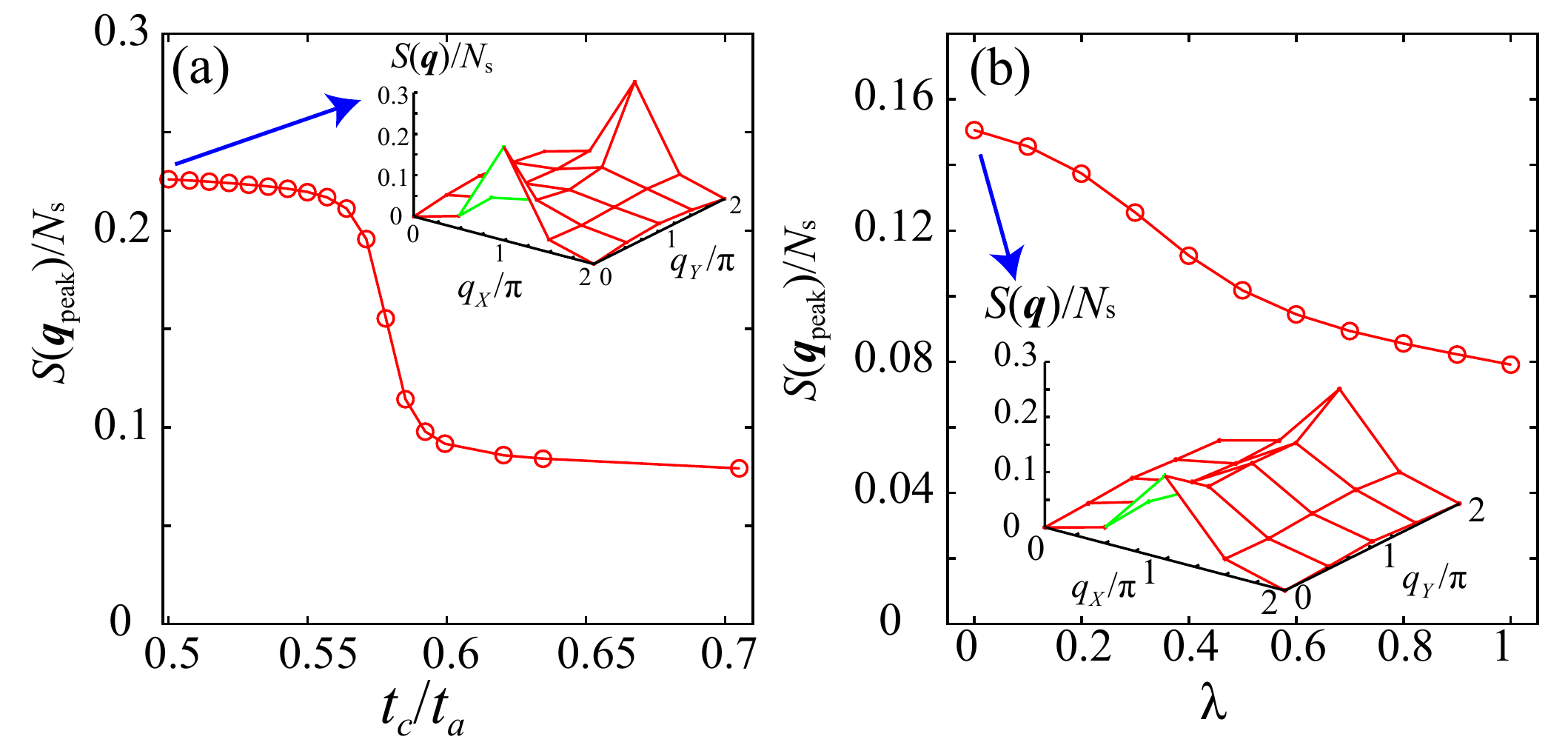}
  \end{center}
\caption{(color online)~
(a) Dependence of $S(\vec{q}_{\rm peak})$ on $t_{c}/t_{a}$.
Around $t_{c}/t_{a}\sim 0.57$, 
the stripe magnetic 
correlations suddenly increase.
The inset shows the spin structure factors
at $t_{c}/t_{a}=0.5$, which indicate the stripe magnetic order. 
(b) Dependence of the peak value of the 
spin structure factors on $\lambda$ for $X$=EtMe$_3$Sb.
The inset shows the spin structure factors at $\lambda=0$.
}
\label{lambda}
\end{figure}

Next, we introduce $\lambda$, which monotonically 
scales the off-site interactions including the
off-site Coulomb interactions ({$V_{ij}$}) and the 
{direct} exchange interactions ({$J_{ij}$}){, i.e., 
($V_{ij},J_{ij}$) is scaled as ($\lambda V_{ij}$,$\lambda J_{ij}$).
}
Note that $\lambda=1$ corresponds to the $ab$ $initio$ Hamiltonian and
$\lambda=0$ corresponds to the simple Hubbard model that has only on-site
Coulomb interactions $U$.
The dependence of 
$S(\vec{q}_{\rm peak})$ on $\lambda$ is shown in Fig. ~\ref{lambda}(b).
$S(\vec{q}_{\rm peak})$ increases by decreasing $\lambda$,  and
{the} 
stripe magnetic order appeared below $\lambda=0.5$ (
the spin structure factor at $\lambda=0$ is shown in the inset).
This result indicates that the off-site interactions, which are
often ignored in the previous studies, 
suppress the magnetic ordered moment
and play an important role in the stabilization of 
the QSL state in Pd(dmit)$_2$ salts.
{We note that both $V_{ij}$ and $J_{ij}$ reduce the magnetic ordered moment
(see S. 5 in \cite{SM})).}
{The mechanism of the reduction 
can be attributed to the reduction of $U$ 
and resultant enhancement of the 
higher-order inteactions such as the ring-exchange interactions,
which suppress the magnetic long-range orders~\cite{Yang_PRL2010,Kenny_PRM2020,Szasz_2018}.
}

{\it Summary---.}
To conclude, 
we have successfully 
reproduced the overall trend in the 
magnetic properties 
by performing comprehensive $ab$ $initio$ calculations
for all available $\beta^{\prime}$-type Pd(dmit)$_2$ salts. 
An antiferromagnetic order with a stripe pattern becomes
the ground state for small $t_{c}/t_{a}$ such as Me$_4$P.
{This result is consistent with experimental results 
that the Neel temperatures increase with decreasing $t_{c}/t_{a}$.}
However, the magnetic ordered moment is 
significantly suppressed around EtMe$_{3}$Sb.
These results indicate that the QSL state
without magnetic order appears in EtMe$_{3}$Sb,
which is consistent with previous experimental results.
{We have also shown that cooperation of the geometrical frustration and
the off-site interactions induce 
suppression of the magnetic ordered moment in EtMe$_{3}$Sb.}
To analyze the QSL state in {molecular solids}, 
most theoretical calculations have been conducted
for 
frustrated Hubbard models that only 
have on-site 
Coulomb interactions~\cite{MoritaPIRG,Tocchio_PRB2009,Jacko_PRB2013}.
The present $ab$ $initio$ calculations, 
which suggest the importance of the off-site interactions,
require reconsideration of the
use of such simple models to 
describe the QSL state in Pd(dmit)$_2$ salts.

Analysis of the low-energy effective model in this Letter is 
limited to small system sizes in order to perform an exact analysis 
to clarify the general trend in the Pd(dmit)$_2$ salts.
Several highly accurate wavefunction methods
for the strongly correlated electron systems
have recently been developed~\cite{Orus_AP2014,becca2017quantum,Misawa_CPC2019}.
{The clarification of 
the possible exotic elementary excitation
of the QSL state in EtMe$_{3}$Sb using such cutting edge methods, 
which is a hotly debated issue 
in experiments~\cite{M_Yamashita_dmit10,BH_PRX2019,Ni_PRL2019,M_Yamashita_PRB2020}, 
is a significant challenge but left for future studies.}
{We also note that the applications of the employed 
$ab$ $initio$ method to other families of the 
molecular solids such as $\kappa$-ET salts
will help us to comprehensively understand the nature of the QSL found in 
other compounds~\cite{Zhou_2017RMP}.}

\begin{acknowledgements}
The calculations were partly conducted at the 
Supercomputer Center, Institute for Solid State Physics, University of Tokyo.
This work was supported by Kakenhi Grants-in-Aid 
(Nos.~JP16H06345, JP19K03739, and 19K21860) from the Japan Society for the Promotion of Science (JSPS).
{The authors thank Kazuma Nakamura for useful discussions on RESPACK.}
{The authors thank Reizo~Kato for stimulating
discussion on the experimental aspects of the materials.}
T.M. and K.Y. thank Hiroshi Shinaoka for a discussion on the cRPA.
T.M. and K.Y. were also supported by the Building of Consortia for 
the Development of Human Resources in Science and Technology from the Ministry of Education, Culture, Sports, Science and Technology (MEXT) of Japan.
T.T. was partially supported by the Leading Initiative for Excellent Young Researchers (LEADER) from MEXT of Japan.
This work was supported by MEXT as ``Program for Promoting Researches on the Supercomputer Fugaku'' 
(Basic Science for Emergence and Functionality in Quantum Matter --Innovative Strongly-Correlated Electron 
Science by Integration of ``Fugaku'' and Frontier Experiments--, Project ID:hp200132). 
\end{acknowledgements}

\clearpage
\noindent
{\Large
Supplemental Material for 
``Electronic Correlation and Geometrical Frustration in
Molecular {Solids} \\
-- A Systematic {\it ab initio} Study of {$\beta^\prime$}-$X$[Pd(dmit)$_{2}$]$_{2}$  ''
}

\section{S.1~Downfolding results}
\begin{figure}[htb!]
  \begin{center}
    \includegraphics[width=7cm,clip]{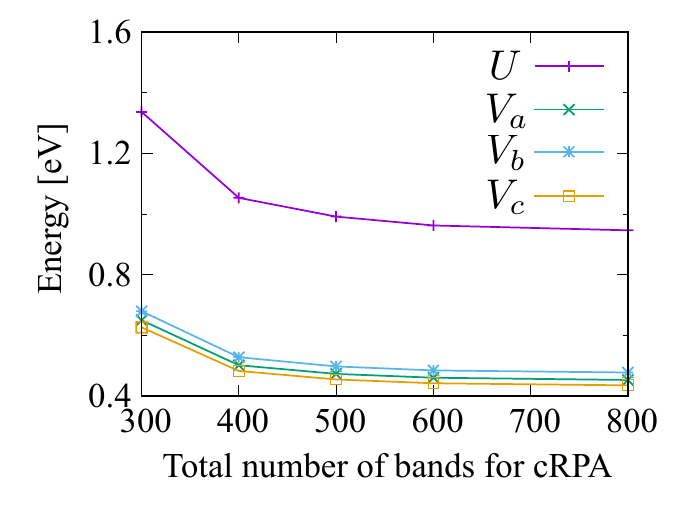}
  \end{center}
\caption{(color online)~
Total number of  bands for cRPA dependence of the on-site and inter-site Coulomb interactions for Et$_2$Me$_2$Sb salt.}
\label{S0}
\end{figure}

The derived interdimer transfer integrals, and both on-site and inter-site Coulomb interaction parameters 
{obtained by constrained Random Phase Approximation (cRPA)}
of $\beta'$-$X$[Pd(dmit)$_{2}$]$_{2}$ are listed in Table.~\ref{TransferUV}. 
The simple transfer integrals reproduce the DFT bands for all the Pd(dmit)$_2$ salts well.  
{In general, the values of on-site and inter-site Coulomb interactions obtained by cRPA depend on the total number of bands $n_b$ contributing to the screening. 
Figure~\ref{S0} shows $n_b$-dependence of on-site and inter-site Coulomb interaction for Et$_2$Me$_2$Sb salt (we select this salt, since it has the largest value of on-site Coulomb interaction in Pd(dmit)$_2$ salts). 
With increasing $n_b$, both on-site and inter-site Coulomb interactions decreases and approach to the constant value. In our manuscript,  $n_b$ is set as $600$.}

\begin{figure}[htb!]
  \begin{center}
    \includegraphics[width=7cm,clip]{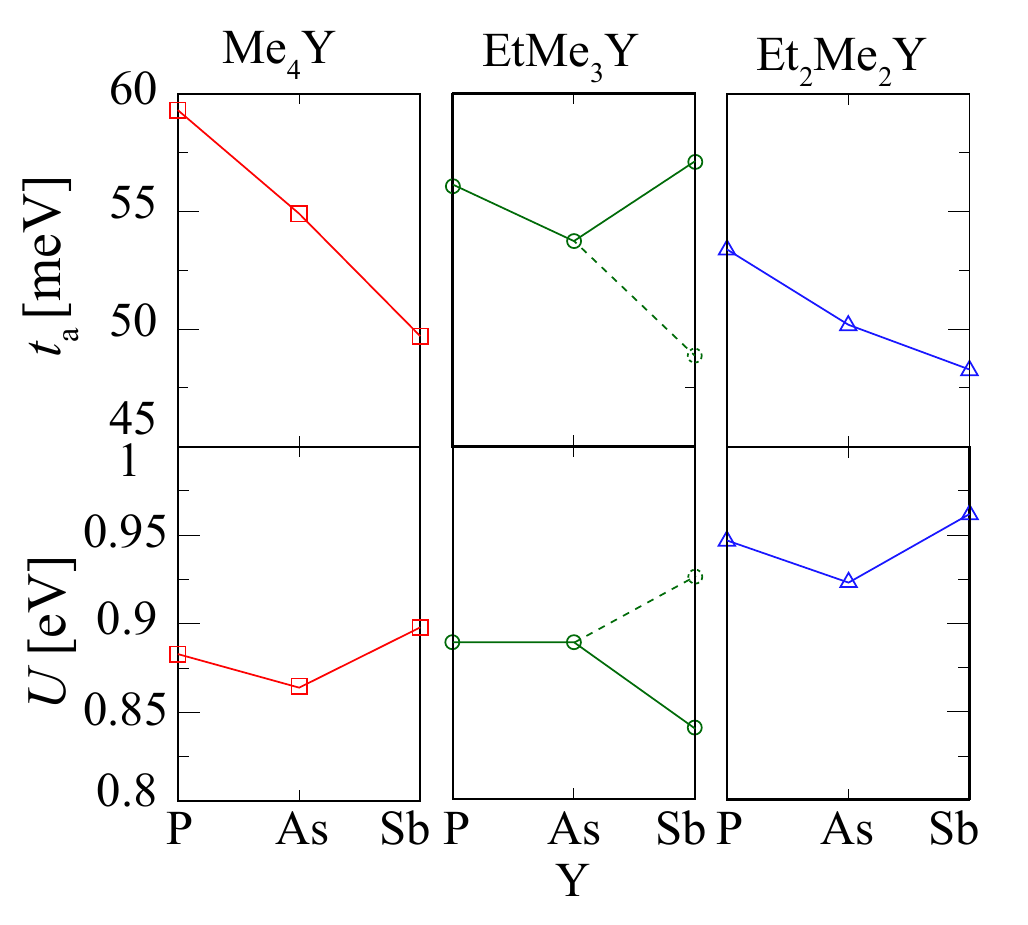}
  \end{center}
\caption{(color online)~
Cation dependence of  the transfer integrals $t_a$ and the on-site Coulomb interactions $U$.
The solid (dotted) symbols for the EtMe$_3$Sb salt show the data calculated 
using the x-ray structure at 4 K (room temperature).}
\label{S1}
\end{figure}

Figure~\ref{S1} shows the cation dependence of the transfer integrals $t_a$, and the on-site Coulomb interactions, $U$. 
The following aspects are noted. Among the parameters for the experimental 
structures measured at room temperature, $t_a$ decreases in the order of P, As, and Sb. 
For the EtMe$_3$Sb salt, 
the distance between dimers, [Pd(dmit)$_2$]$_2$, becomes shorter with decreasing temperature, and thus $t_a$ become larger, and the anisotropy among three transfer integrals increases.  
In contrast, the cation dependence of the on-site interaction $U$ is 
weak compared to that of $t_a$.

Experimentally, the EtMe$_3$Sb salt does not show the phase transition down to 
the lowest temperature {($\sim$ 32 mK)}, 
while the other compounds show phase transitions 
such as the antiferromagnetic transition and the charge ordering transition. 
To obtain the effective model with no ordered patterns, 
we adopt the crystal data of EtMe$_3$Sb salt at 4 K for the downfolding calculation, 
while for the other compounds, we adopt those obtained at room temperature. 

\begin{table*}[htb]
  \begin{tabular}{lccccccccccc} \hline
    Cation & Temperature & $t_a$[meV] & $t_b$ [meV]  & $t_c$ [meV] & $U$ [eV] & $V_a$ [eV] & $V_b$ [eV] & $V_c$ [eV] & $J_a$[meV] & $J_b$ [meV]  & $J_c$ [meV] \\ \hline
    Me$_4$P & rt  &59.3 & 44.0 & 30.9&0.883 & 0.449 & 0.465 & 0.413 &3.26&2.54&1.29\\
    Me$_4$As &rt  & 54.9 & 41.5 & 33.8 & 0.864 & 0.426 & 0.442 & 0.394&2.77&2.37	&1.35\\
    Me$_4$Sb & rt  &49.7 & 36.5 & 38.9& 0.898 & 0.429 & 0.444 & 0.402&2.67&1.75&1.49\\ \hline
    EtMe$_3$P & rt &56.1 & 42.0 & 35.0& 0.889 &0.442& 0.460 &0.411&2.88&2.02&1.48\\
    EtMe$_3$As & rt &53.7 & 40.9 &37.2&  0.889 & 0.436 & 0.456 & 0.409&2.60&1.84&1.45\\
    EtMe$_3$Sb & rt & 48.8& 35.6 & 41.7& 0.906 &  0.427& 0.449 & 0.406&3.21&2.23&1.40\\ 
    EtMe$_3$Sb & 4K & 57.1& 44.6 & 40.3& 0.840 & 0.413 & 0.434 & 0.390&2.23&1.64&1.71\\ \hline
    Et$_2$Me$_2$P & rt &53.4 & 38.5 & 38.6& 0.947 & 0.478 & 0.497 & 0.450&2.51&2.32&1.43\\
    Et$_2$Me$_2$As & rt &50.2 & 36.3 & 38.7& 0.923 &0.422 & 0.467 & 0.448&3.30&2.63&1.73\\
    Et$_2$Me$_2$Sb & rt &48.3 & 33.5 & 45.3 & 0.962 &0.461 & 0.485 & 0.443&2.41&1.77& 1.58\\ \hline
  \end{tabular}
\caption{List of the parameters obtained by the downfolding method in the dimer-model extended Hubbard-type Hamiltonian for $\beta'$-$X$[Pd(dmit)$_{2}$]$_{2}$ (rt represents room temperature).}
\label{TransferUV}
\end{table*}

\section{S.2~Dimensional downfolding}
Here, we explain the 
effects of dimensional downfolding.
The models that have three-dimensional 
interactions are obtained by performing constrained random phase approximation (cRPA). To reflect the two-dimensionality 
of the organic solids and reduce numerical costs, 
it is better to directly treat
the two-dimensional model.
To do so, Nakamura $et$ $al$.  proposed a
dimensional downfolding method~\cite{Nakamura_JPSJ2010,Nakamura},
which renormalizes the three-dimensional
interactions into two-dimensional interactions.
As a result, it is shown that
the dimensional downfolding simply induces
a constant shift of the Coulomb interaction, irrespective of
the distance, i.e.,
\begin{align}
\tilde{V}_{ij}=V_{ij}-\Delta.
\end{align}
The amplitudes of the constant shifts are 
evaluated as $\Delta=0.18$ eV for the dmit salt (EtMe$_3$Sb)
and $\Delta=0.2$ eV for {$\kappa$-(BEDT-TTF)} salt~\cite{Nakamura}.
{
For $\Delta\geq0.3$, $\tilde{V}_{ij}$ for [-11] direction
becomes negative for several compounds.
In those cases, we simply ignore them, i.e., 
we set the small but finite $\tilde{V}_{ij}<0$ as 0. }

\begin{figure}[htb!]
  \begin{center}
    \includegraphics[width=9cm,clip]{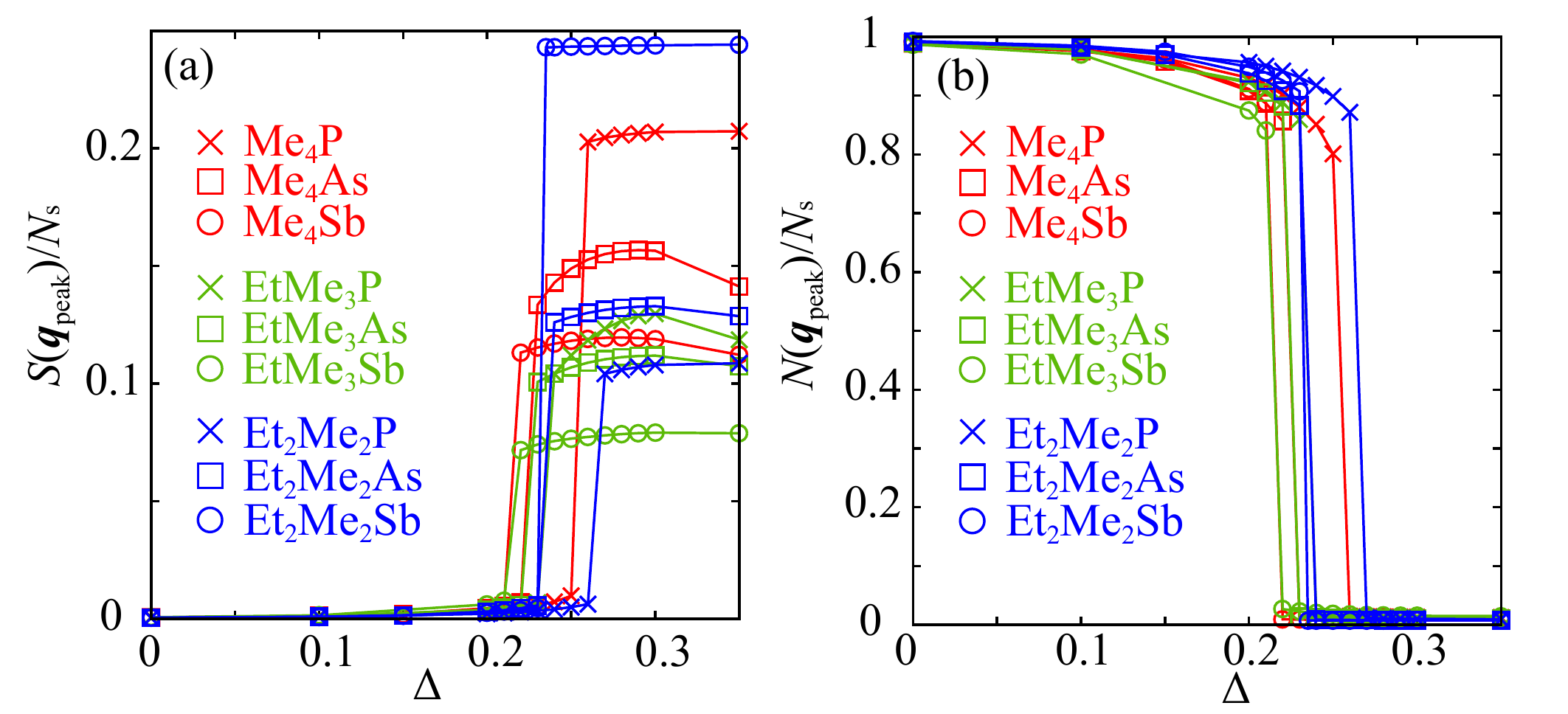}
  \end{center}
\caption{(color online)~
(a) Dependence of the spin structure factors on $\Delta$
for 9 compounds.
(b) Dependence of the charge structure factors on $\Delta$
for 9 compounds.
}
\label{S2}
\end{figure}

\begin{figure}[htb!]
  \begin{center}
    \includegraphics[width=7cm,clip]{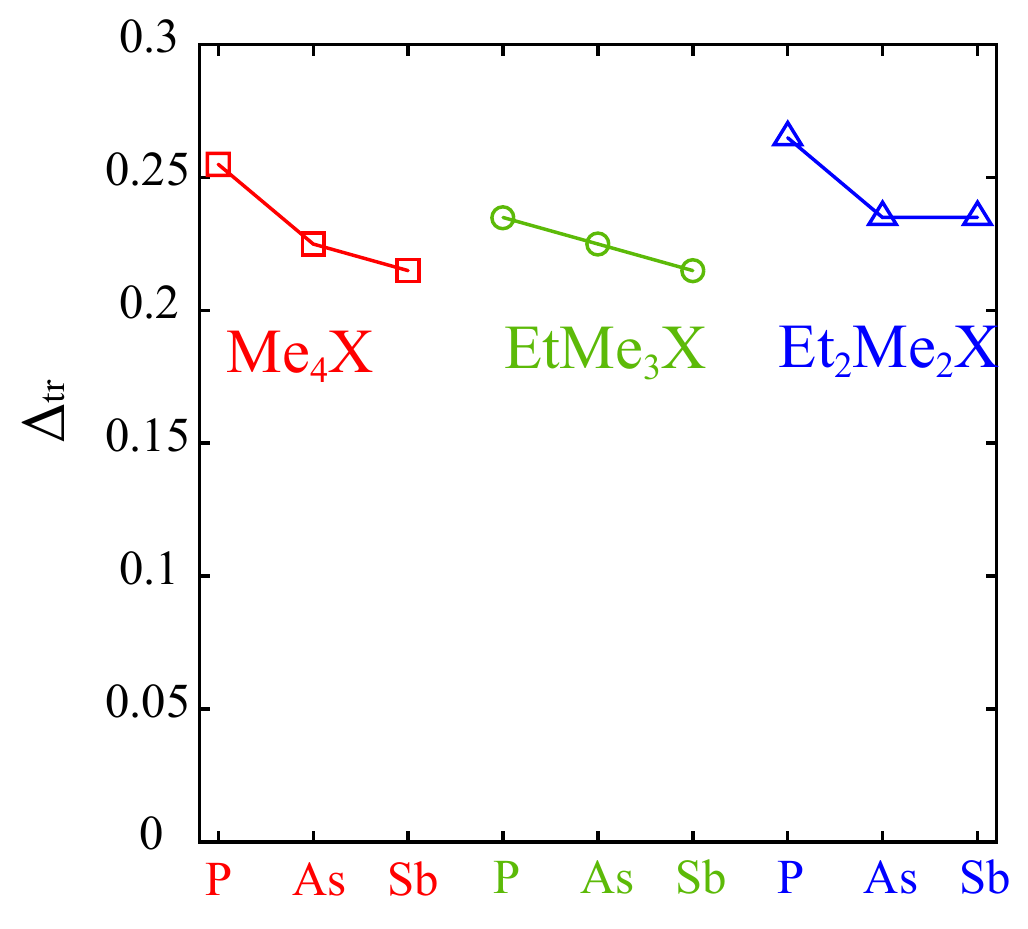}
  \end{center}
\caption{(color online)~
Compound dependence of the transition value of $\Delta$ ($\Delta_{\rm tr}$). 
}
\label{S3}
\end{figure}

We examine the effects of 
the dimensional downfolding, i.e., how the 
constant shift $\Delta$ affects the electronic structures in
the low-energy effective model.
By changing $\Delta$, the effective models are solved
using exact diagonalization~\cite{hphi}.
In Fig.~\ref{S2}, we show the dependence of 
the spin and charge structure factors on $\Delta$,
which are defined as
\begin{align}
S(\vec{q}) &= \frac{1}{N_{\rm s}}\sum_{i,j}\langle\vec{S}_{i}\cdot\vec{S}_{j}\rangle {\rm e}^{i\vec{q}\cdot(\vec{r}_{i}-\vec{r}_{j})} \\
N(\vec{q}) &= 
\frac{1}{N_{\rm s}}\sum_{i,j}\langle({N}_{i}-\langle N_{i}\rangle)({N}_{j}-\langle N_{j}\rangle)\rangle {\rm e}^{i\vec{q}\cdot(\vec{r}_{i}-\vec{r}_{j})}
\end{align}
The charge order state
becomes the ground state for $\Delta=0$
for all 9 compounds (that is, the ordering vector for the charge ordered state is $(\pi,0)$).
As $\Delta$ is increased, 
a discontinuous phase transition occurs
between the charge order state and the antiferromagnetic
states around $\Delta_{\rm tr}=0.2-0.26$ eV.
The compound dependence of the transition points
is shown in Fig.~\ref{S3}.

To 
{perform} a parameter-free approach and
examine the magnetic properties 
of the Pd(dmit)$_2$ salts,
$\Delta=0.3$ was employed in this paper. Although this value is slightly larger than that
obtained by previous studies,
the magnetic properties
are not largely dependent on $\Delta$ 
{in the magnetic ordered phase,}
as shown in Fig.~\ref{S2}(a).
Therefore, this constant shift does not significantly 
change the result.

\clearpage
\section{S.3~Compound dependence of spin structure factors}
\begin{figure}[h!]
  \begin{center}
    \includegraphics[width=14cm,clip]{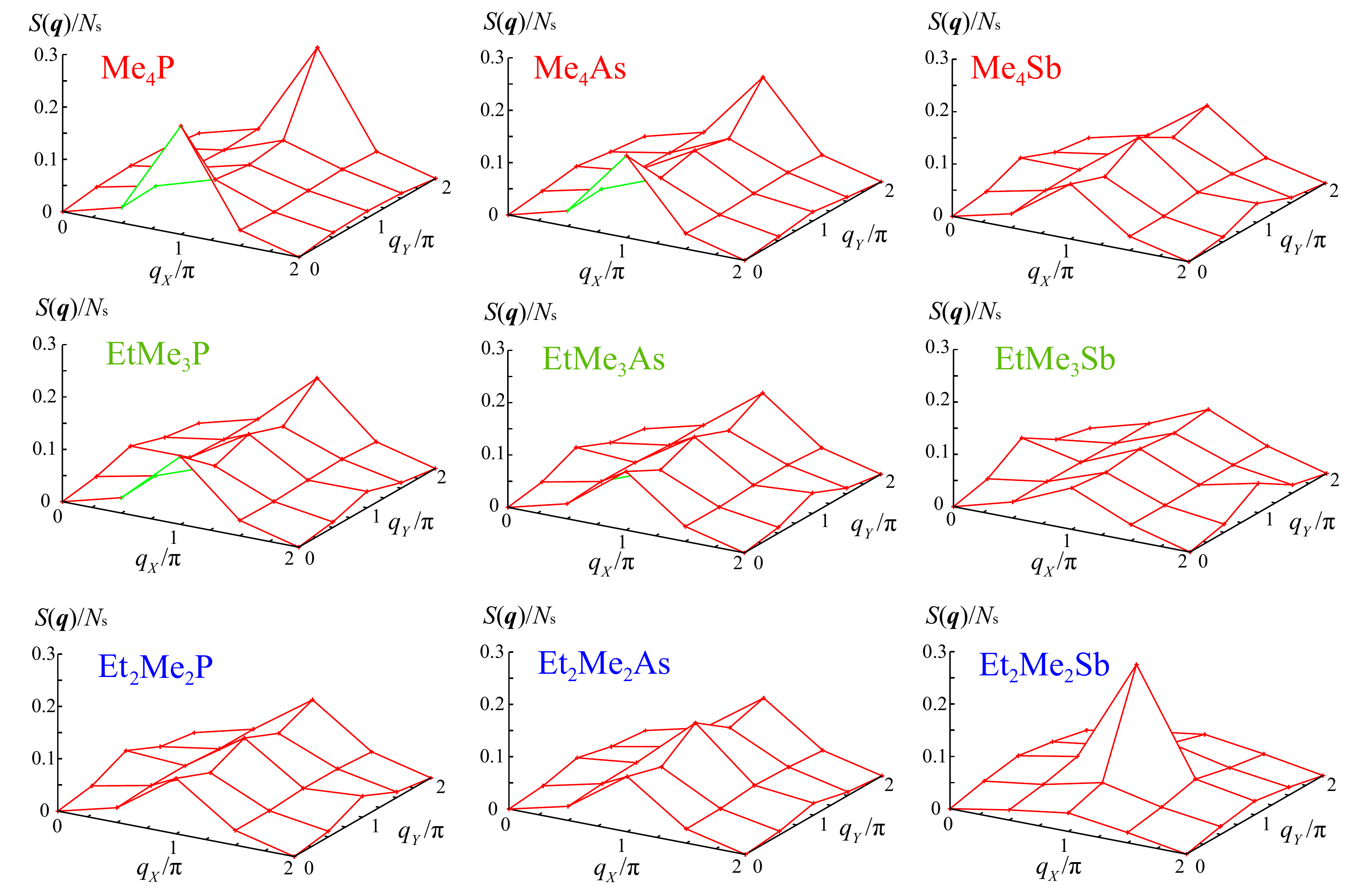}
  \end{center}
\caption{(color online)~
Compound dependence of the spin structure factors.
}
\label{S4}
\end{figure}

Although the spin structure factors for three typical compounds
are shown in the main text, for comparison, we show the spin structures 
for all 9 compounds in Fig.~\ref{S4}.
For Me$_4$P, the stripe magnetic correlations 
are dominant. This stripe order is suppressed by 
changing the cation, i.e., by increasing $t_{c}/t_{a}$, and
a non-magnetic phase is apparent around EtMe$_3$Sb.
For Et$_2$Me$_2$Sb, the Neel magnetic correlation becomes
dominant because 
it is close to the square lattice $t_{c}/t_{a}\sim 0.8$.

\clearpage
\section{S.4~Bond-order structure factors}
\begin{figure}[h!]
  \begin{center}
    \includegraphics[width=6cm,clip]{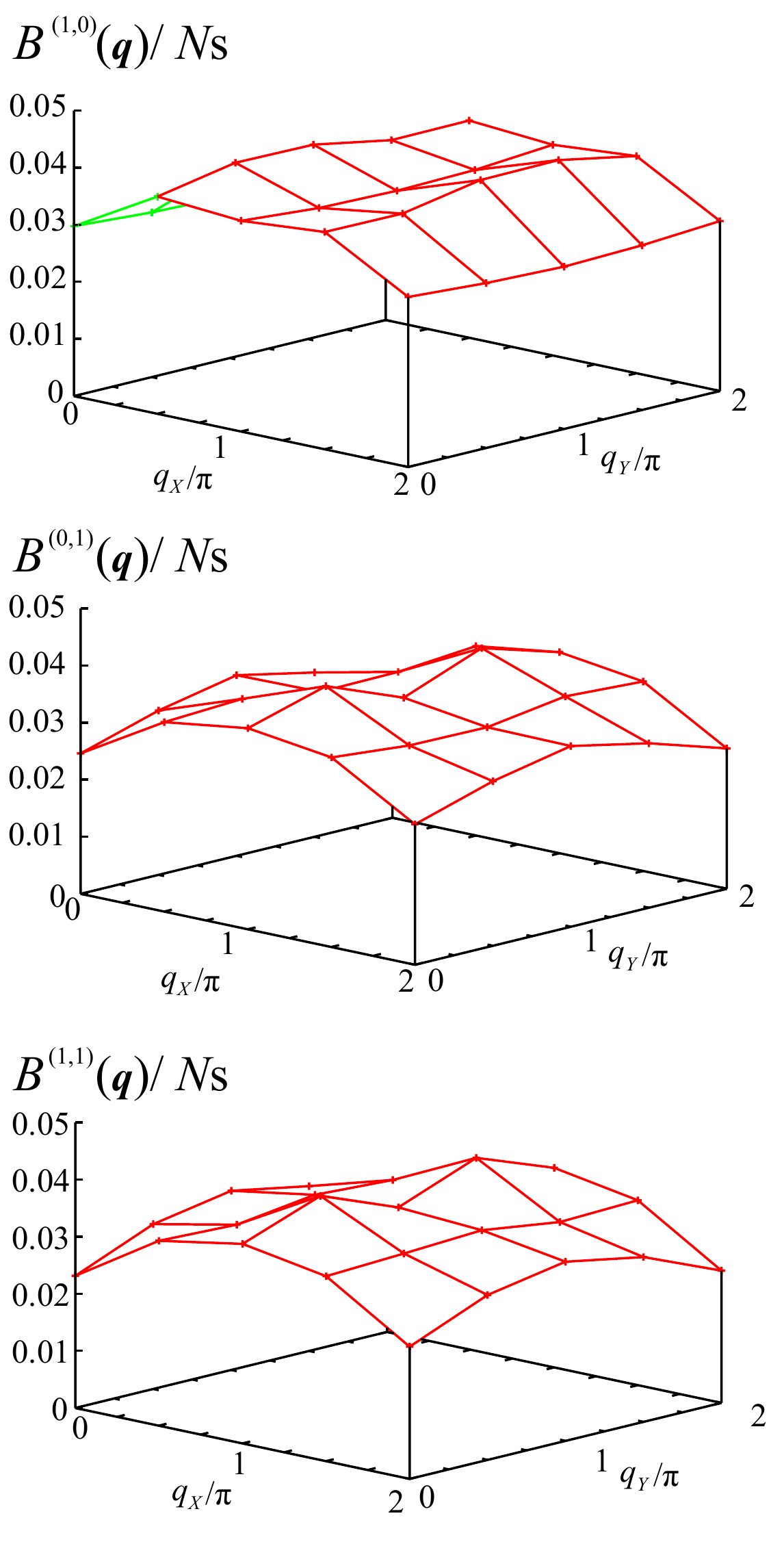}
  \end{center}
\caption{(color online)~
Bond-order structure factors for 
the low-energy effective model of EtMe$_3$Sb.
We take $\vec{e}=(1,0),(0,1)$, and $(1,1)$.
}
\label{S5}
\end{figure}

One of the possible candidates
for a non-magnetic phase in the extended Hubbard
model is the bond-order phase, which becomes a
ground state in the one-dimensional extended 
Hubbard model~\cite{MNakamura_JPSJ1999,MNakamura_PRB2000}.
The bond-order structure factors were calculated
for the low-energy effective model of EtMe$_{3}$Sb,
which is defined as
\begin{align}
B^{\vec{e}}(\vec{q}) &= 
\frac{1}{N_{\rm s}}\sum_{i,j}
\langle({B}_{i}^{\vec{e}}-\langle {B}_{i}^{\vec{e}}\rangle)
({B}_{j}^{\vec{e}}-\langle {B}_{j}^{\vec{e}}\rangle)
\rangle {\rm e}^{i\vec{q}\cdot(\vec{r}_{i}-\vec{r}_{j})} \\
B_{i}^{\vec{e}} &= \frac{1}{2}
\sum_{\sigma}(c_{i+\vec{e},\sigma}^{\dagger}c_{i,\sigma}
+c_{i,\sigma}^{\dagger}c_{i+\vec{e},\sigma}),
\end{align}
where $\vec{e}$ represents the direction of the bond-order.
Figure~\ref{S5} shows the bond-order
structure factors in the low-energy effective model for EtMe$_3$Sb.
No clear signature of the bond order was evident.

\clearpage
\section{S.5~Effects of off-site interactions}
\begin{figure}[h!]
  \begin{center}
    \includegraphics[width=9cm,clip]{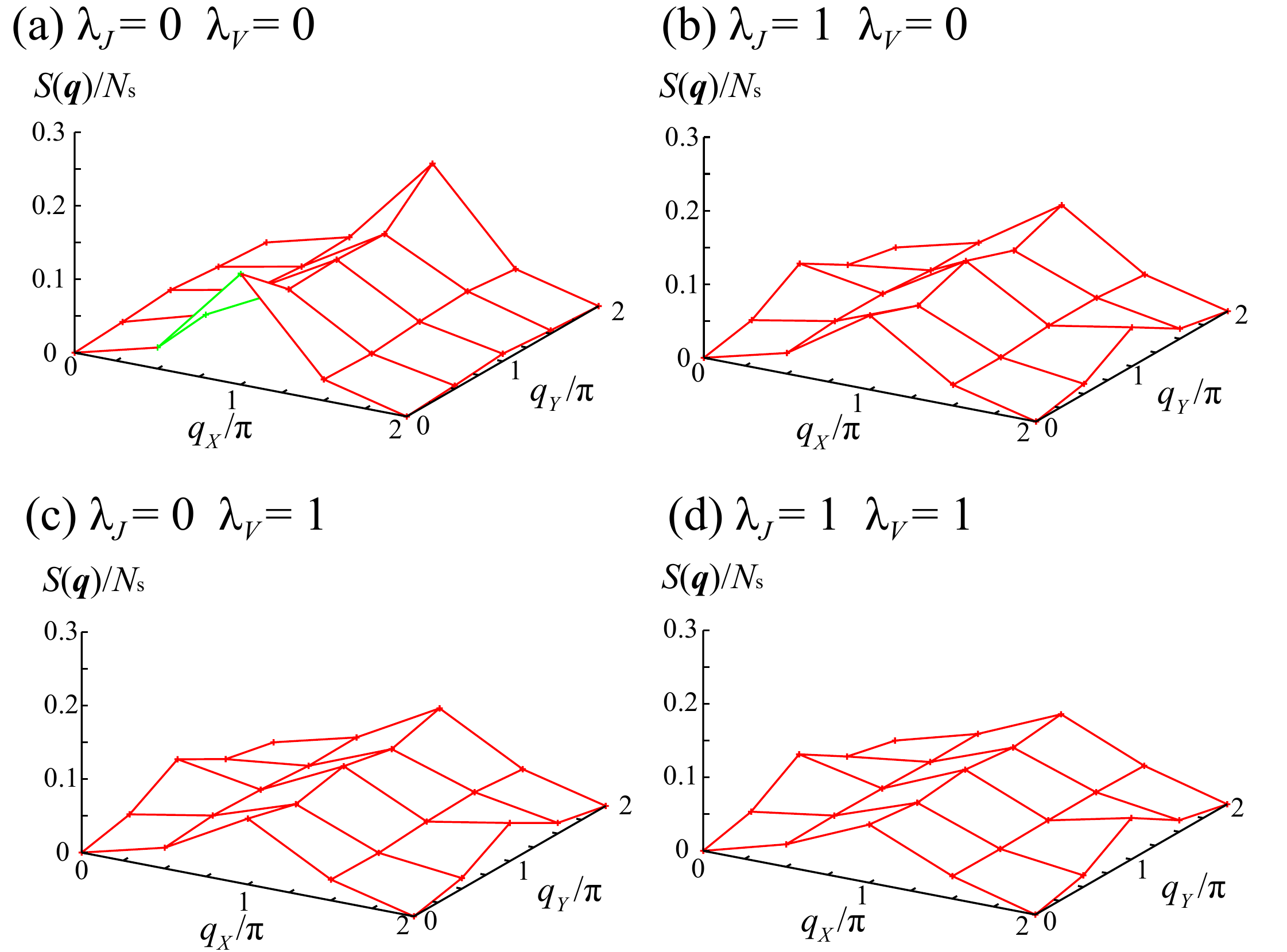}
  \end{center}
\caption{(color online)~
Spin structures for 
(a)~$\lambda_J=\lambda_V=0$ (only on-site Coulomb interaction $U$), 
(b)~$\lambda_J=1$, $\lambda_V=0$,
(c)~$\lambda_J=0$, $\lambda_V=1$, and
(d)~$\lambda_J=\lambda_V=1$ ($ab$ $initio$ Hamiltonian).
Both the off-site Coulomb interactions and
direct exchange interactions reduce the magnetic ordered moment.
}
\label{S6}
\end{figure}

{
To examine the effects of the off-site Coulomb interactions
and the direct exchange interactions separately,
we introduce the scaling parameters $\lambda_V$ and $\lambda_J$, which scale 
($V_{ij},J_{ij}$) as ($\lambda_V V_{ij}$,$\lambda_J J_{ij}$).
We calculate the spin structure factors of the
Hamiltonians for $\lambda_V=0,1$ and $\lambda_J=0,1$.
As shown in Fig.~\ref{S6},
we find that
both the off-site Coulomb interactions and
the direct exchange interactions significantly reduce the
magnetic ordered moment.
}

\clearpage
\providecommand{\noopsort}[1]{}\providecommand{\singleletter}[1]{#1}%

\end{document}